# Temporal credit assignment for one-shot learning utilizing a phase transition material

Alessandro R. Galloni*,[1], Yifan Yuan*,[2], Minning Zhu[2], Haoming Yu[3], Ravindra S. Bisht[2], Chung-Tse Michael Wu[2], Christine Grienberger[4], Shriram Ramanathan[†,2] and Aaron D. Milstein[†,1]

[1] Department of Neuroscience and Cell Biology, Robert Wood Johnson Medical School and Center for Advanced Biotechnology and Medicine, Rutgers, The State University of New Jersey, Piscataway, NJ
[2] Department of Electrical and Computer Engineering, Rutgers, The State University of New Jersey, Piscataway, NJ
[3] School of Materials Engineering, Purdue University, West Lafayette, IN
[4] Department of Neuroscience, Department of Biology and Volen National Center for Complex Systems, Brandeis University, Waltham, MA

* These authors contributed equally to this work.
[†] Co-corresponding authors.
Please send correspondence to:
S.R. (shriram.ramanathan@rutgers.edu)
A.D.M. (milstein@cabm.rutgers.edu)

## Abstract
Design of hardware based on biological principles of neuronal computation and plasticity in the brain is a leading approach to realizing energy- and sample-efficient artificial intelligence and learning machines. An important factor in selection of the hardware building blocks is the identification of candidate materials with physical properties suitable to emulate the large dynamic ranges and varied timescales of neuronal signaling. Previous work has shown that the all-or-none spiking behavior of neurons can be mimicked by threshold switches utilizing phase transitions. Here we demonstrate that devices based on a prototypical metal-insulator-transition material, vanadium dioxide ($VO_2$), can be dynamically controlled to access a continuum of intermediate resistance states. Furthermore, the timescale of their intrinsic relaxation can be configured to match a range of biologically-relevant timescales from milliseconds to seconds. We exploit these device properties to emulate three aspects of neuronal analog computation: fast (~1 ms) spiking in a neuronal soma compartment, slow (~100 ms) spiking in a dendritic compartment, and ultraslow (~1 s) biochemical signaling involved in temporal credit assignment for a recently discovered biological mechanism of one-shot learning. Simulations show that an artificial neural network using properties of $VO_2$ devices to control an agent navigating a spatial environment can learn an efficient path to a reward in up to 4-fold fewer trials than standard methods. The phase relaxations described in our study may be engineered in a variety of materials, and can be controlled by thermal, electrical, or optical stimuli, suggesting further opportunities to emulate biological learning.

**Main**

Adaptive behavior in biological organisms and engineered systems depends critically on experience-dependent learning. In the brain, neuronal plasticity mechanisms store information and enable recall of past experiences. Insight into the mechanisms of biological neuronal information processing and plasticity has inspired the design of modern artificial neural networks (ANNs), which can be trained to perform classification and prediction tasks with superhuman accuracy[1]. However, while biological neuronal networks in the human brain are capable of learning based on few or even single experiences[2], the standard supervised and reinforcement-based algorithms used to train ANNs are inefficient, often requiring repeated sampling of thousands to billions of data samples and utilizing millions of compute hours[3]. Learning to make predictions from temporal sequences of inputs has particularly high energy demands using standard machine learning approaches. The recent history of activity of all neurons in the network in response to past sequence elements must be stored in memory and consulted to adjust synaptic connection strengths (weights) and improve performance[4,5]. These requirements pose a major challenge to the design of energy-efficient learning machines that could adapt to real-world experience after initial training and deployment[6]. Thus it is a shared goal of research in machine learning and computational neuroscience to identify the biological synaptic plasticity rules, neural circuit architectures, and learning algorithms that enable the brain to rapidly adapt, and to engineer ANNs that utilize those principles to learn more efficiently[7-9]. Neuromorphic computing aims to further translate insight from biological mechanisms of computation, learning, and memory to the design of energy-efficient hardware[10-15].

One important insight is that connected pairs of biological neurons communicate via brief, infrequent, discrete signaling events called action potentials. Emulating this temporally and spatially sparse signaling behavior can reduce the energy and memory requirements of an artificial network by eliminating the need to store and broadcast the activities of all neurons at each training step[16]. Biological neurons are also well-suited for discriminating temporal sequences of inputs, as they express multiple ion channels and signaling proteins that act as intermediate state variables and store information about recent history of activation across multiple timescales from milliseconds to seconds[17-20]. This suggests that incorporating analog internal state variables that change on biologically-relevant timescales into hardware implementations of neurons could improve performance on sequence discrimination tasks while also reducing memory and data transport requirements[12,21,22].

Another line of work suggests that local storage of slow neuronal state variables can enable powerful one-shot learning mechanisms. Recently, a novel biological learning rule was characterized in rodent hippocampal neurons that is gated by slow calcium spikes in neuronal dendrite compartments, and stores associative memories based on a single experience[23-25]. This behavioral timescale synaptic plasticity (BTSP) utilizes seconds-long intracellular biochemical traces to associate events such as visual cues and rewards across long time delays[23-25]. According to the BTSP learning rule, presynaptic spikes increment a slowly decaying biochemical "eligibility trace" (ET) that marks recently active synapses as eligible for synaptic plasticity for an extended time period[20,22-24]. This assigns credit to synapses that are active within a seconds-long time window leading up to a behavioral outcome like receipt of reward[24,25]. Then, when a postsynaptic neuron receives a supervisory input to its distal dendrites, it emits a dendritic calcium spike. This, in turn, activates another slowly decaying biochemical "instructive signal" (IS) that is required for synaptic plasticity and modifies the strengths of all synapses that are tagged with nonzero ETs[23-25].

     Here, we designed and simulated a neuromorphic circuit to emulate these temporal integration and associative memory properties of biological neurons. We show that the resistance of a prototypical metal-insulator transition material, vanadium dioxide ($VO_2$)[26], can be dynamically controlled to mimic neuronal internal state variables, enabling temporal credit assignment for one-shot learning with the BTSP learning rule (Fig. 1a-c). These results demonstrate that incorporating emerging materials and engineering their unique properties can contribute to the design of biology-inspired computing devices.

**Metal-insulator phase transitions at biological timescales**
$VO_2$ is an exemplary correlated electron quantum material that is insulating at room temperature, but undergoes a structural change to a metallic phase that is highly conductive when exposed to elevated temperature or electrical stimulation (Fig. 1a,b)[27,28]. Much interest in $VO_2$ across the natural sciences and engineering disciplines is due to the ultra-fast nature of the phase transition that can happen even at ~$10^{-15}$ second timescales[29]. At the same time, the transition voltage and timescale can be controlled by choice of the ambient temperature, driving stimulus, circuitry, sample geometry, and chemical doping, opening up new frontiers in emerging computing[30-39]. Here we focused specifically on biological timescales (~1 millisecond to 10 seconds) that correspond to different neuronal intracellular electrical and biochemical signals. We fabricated an array of $VO_2$ devices in which high-quality epitaxial $VO_2$ thin films are grown on sapphire substrates (Fig. 2a). The structure and chemical composition of $VO_2$ thin films were characterized by X-ray diffraction and Rutherford backscattering spectrometry (RBS) (Supplementary Fig. S1 and S2; see Supplementary Information). The device conductance was measured while varying temperature and applied voltage (Fig. 2b and Supplementary Fig. S1b). When the amplitude of applied voltage is slowly ramped up, the initially low conductance of the material increases (Fig. 2b, red), with a rapid

**Fig. 1.**

**a** Controllable conductance and relaxation timescales of VO₂ devices

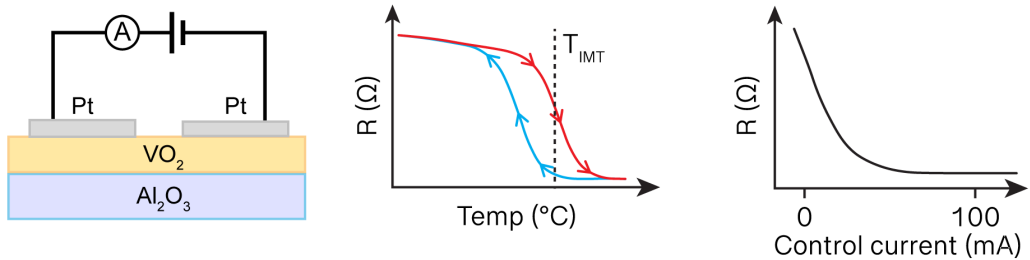

**b** Neuromorphic analog computing

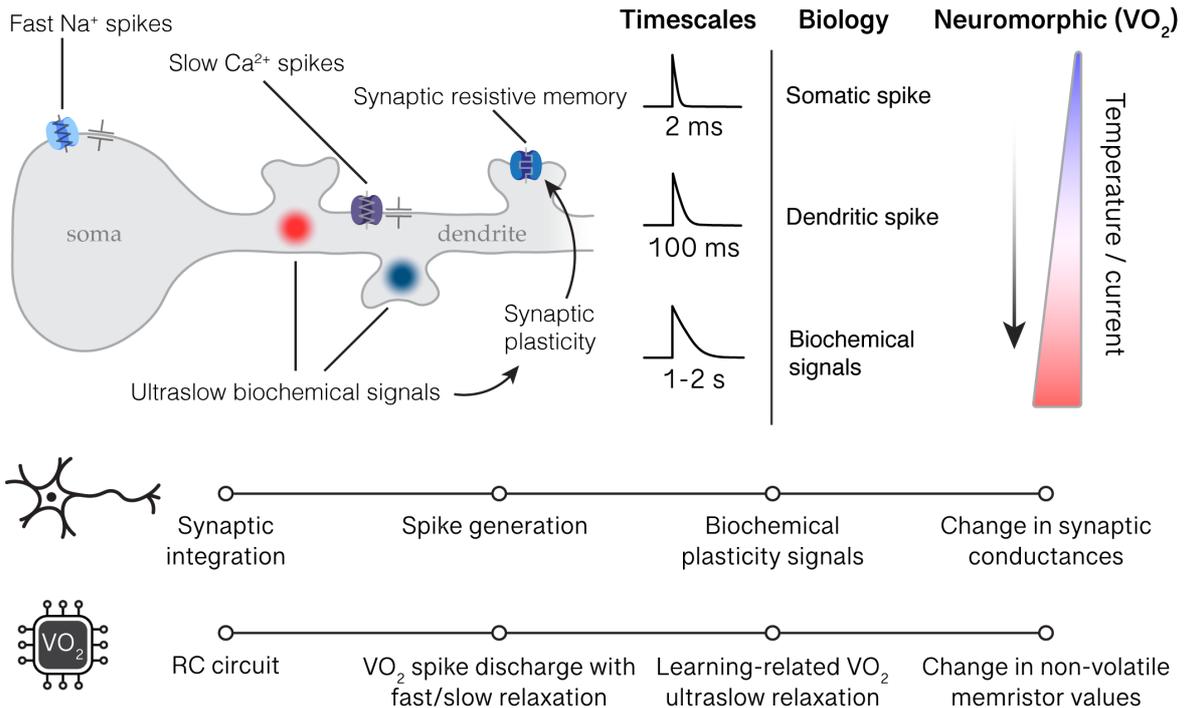

**c** One-shot reinforcement learning

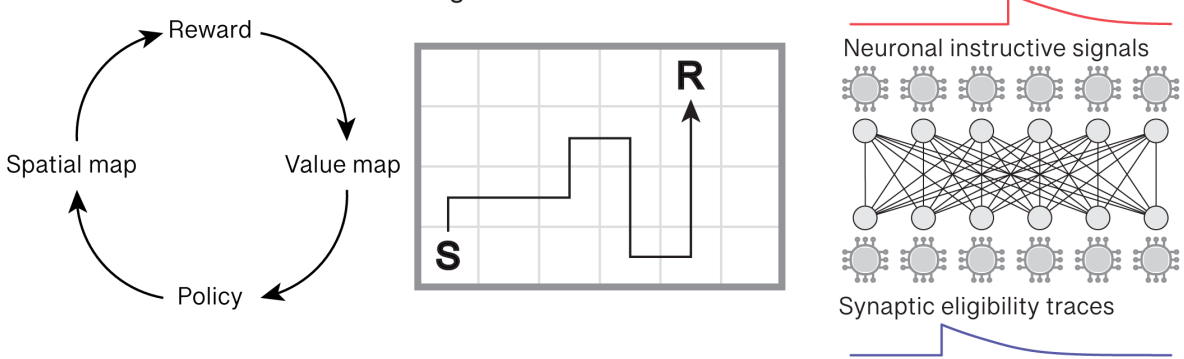

**Fig. 1. Neuromorphic analog computing at biological timescales with a phase transition material.**
**a**, Left: Schematic of vanadium oxide (VO₂)-based device. Middle: Illustration of relationship between device resistance and temperature during heating (red) and cooling (blue). $T_{IMT}$ indicates insulator-metal-transition threshold temperature. Right: Illustration of relationship between device resistance and applied current. **b**, Diagram depicts electrical and biochemical signaling events involved in synaptic integration and plasticity in neurons.

Voltage-gated sodium and potassium channels in the cell body (soma) of a neuron mediate fast action potentials (spikes) with a short duration and refractory period (~2 ms). Voltage-gated calcium channels and calcium-sensitive potassium channels in dendritic compartments of neurons mediate slow calcium spikes with a long duration and refractory period (~100 ms). The amplitudes of synaptic currents control the stimulus selectivity and spiking rates of neurons, and can be modified by learning. Non-volatile resistive memory elements (memristors) are commonly used in neuromorphic computing to represent modifiable synapses for on-chip learning. Intracellular biochemical signals that mediate synaptic plasticity (e.g. protein kinases) decay at slower (1-2 s) timescales. In this study, we show that the variable resistance and intrinsic relaxation of the phase transition material $VO_2$ can be modulated by temperature and electrical stimulation to match these biological timescales. **c**, Diagrams depict use of $VO_2$ phase relaxations to perform temporal credit assignment for synaptic plasticity in a neuromorphic network during spatial navigation reinforcement learning. Left: As a virtual agent explores a spatial environment (center), a spatial map is learned through experience-dependent plasticity. Reward encounters update a map of the expected value of positions in the environment. This value map is consulted to determine the action policy of the agent, which in turn biases the experience of the agent and re-shapes the spatial map. Center: a two-dimensional grid of spatial positions with the start (S) and reward (R) locations marked. Right: Diagram depicts a neuromorphic network of neuronal units. For each presynaptic input neuron, a $VO_2$-based variable resistor is used to represent a slowly decaying internal state variable that marks a synaptic input as recently active and eligible to undergo synaptic plasticity. For each postsynaptic output neuron, a $VO_2$ device is used to represent a distinct slowly decaying internal state variable that instructs all synapses with nonzero eligibility to undergo plasticity. This biology-based synaptic plasticity rule enables one-shot learning of temporally-extended spatial receptive fields that store an efficient path to reward in a single trial.

transition to high conductance at a critical voltage threshold. Then, as the amplitude of the applied voltage is slowly ramped down, the device transitions back from high to low conductance (Fig. 2b, blue). We next tested if the large dynamic range of intermediate resistance states could be accessed by applying voltage steps to the device at varying points during the phase relaxation of the material. Indeed, we found that voltage steps applied when the material was in arbitrary intermediate resistance states could rapidly and stably transition the material to different resistance states depending on the amplitude of the applied voltage (Fig. 2c). We noted during this experiment that when the applied voltage was reduced to a negligible amplitude to measure the resistance of the material without stimulating it, the resistance intrinsically relaxed to its maximum value over a time period of multiple seconds (Fig. 2c).

We next examined if the relaxation timescale of the device could be modified to be either faster or slower to match the range of timescales characteristic of biological intracellular signals. To test this possibility, we activated intermediate conductance states of the material with transient pulses of electrical stimulation (1 s in Fig. 2d and 10 ms in Fig. 2e), and then measured the relaxation time constant of the material's conductance after the stimulus pulse had ceased (Fig. 2f). Under these conditions, we found that varying either the ambient temperature or the amplitude of the transient activating stimulus pulse resulted in orders of magnitude changes in the relaxation timescale of the material's conductance (Fig. 2d-f). This showed that devices can be configured such that the conductance of the material intrinsically decays with fast (~1 ms), slow (~100 ms), or ultraslow (~500 ms) timescales (Fig. 2d,f). Strikingly, these timescales match those of three important biological neuronal signaling elements: fast (~1 ms) voltage-gated sodium and potassium channels that mediate somatic spikes, slow (~100 ms) voltage-gated calcium channels and calcium-activated potassium channels that mediate dendritic calcium spikes[40-42], and the ultraslow (>500 ms) biochemical traces that underlie BTSP[23,24]. We next sought to design and simulate a neuromorphic circuit to exploit these material properties for analog computing and associative learning. In order to simulate the essential material properties, we developed a parameterized mathematical model of $VO_2$ conductance dynamics and fit the model to mimic the hardware experimental measurements (Fig. 2g-i and Supplementary Fig. S3; see Methods).

**Design of a scalable hybrid circuit for neuromorphic computing**

Hybrid electrical circuits that integrate emerging materials with traditional silicon-based complementary metal-oxide semiconductor (CMOS) components are an active area of research in neuromorphic computing and have been shown to emulate complex behaviors and timescales of neurons while utilizing fewer circuit elements, reduced circuit area, and less energy to operate compared to equivalent pure CMOS-solutions[39,43-46]. Materials that exhibit resistive switching, including metal oxides like $VO_2$ and niobium dioxide, metal sulfides, and inorganic insulators have been successfully deposited at nanoscale onto CMOS-compatible substrates and used to emulate the integrative and spiking properties of neurons[34,39,43-45,47-49]. Previous work has shown that when $VO_2$ devices are scaled down in size (~100 nm – 5 μm), the voltage and current required to induce the insulator-metal-transition (IMT) are significantly reduced (<1 V and <1 mA)[32,33,43,48-50]. To verify this, we measured the threshold current required to induce IMT, and compared devices fabricated with different distances between electrodes. While a $VO_2$ device with a 100 μm gap (Fig. 2) required >1 mA to induce IMT, a device with a 2.5 μm gap required only <0.1 mA (Supplementary Fig. S4a). We also confirmed that slow relaxations occur at elevated temperature in the smaller device (Supplementary Fig. S4c).

In our experiments and model (Fig. 2d,f,h,i and Supplementary Fig. S3), we parameterized ambient temperature to configure the relaxation timescale of $VO_2$ devices. We note that it is not necessary to thermally isolate individual neighboring devices from each other on a two-dimensional array. When an activating current is applied to a small region of material, Joule heating locally increases the temperature,

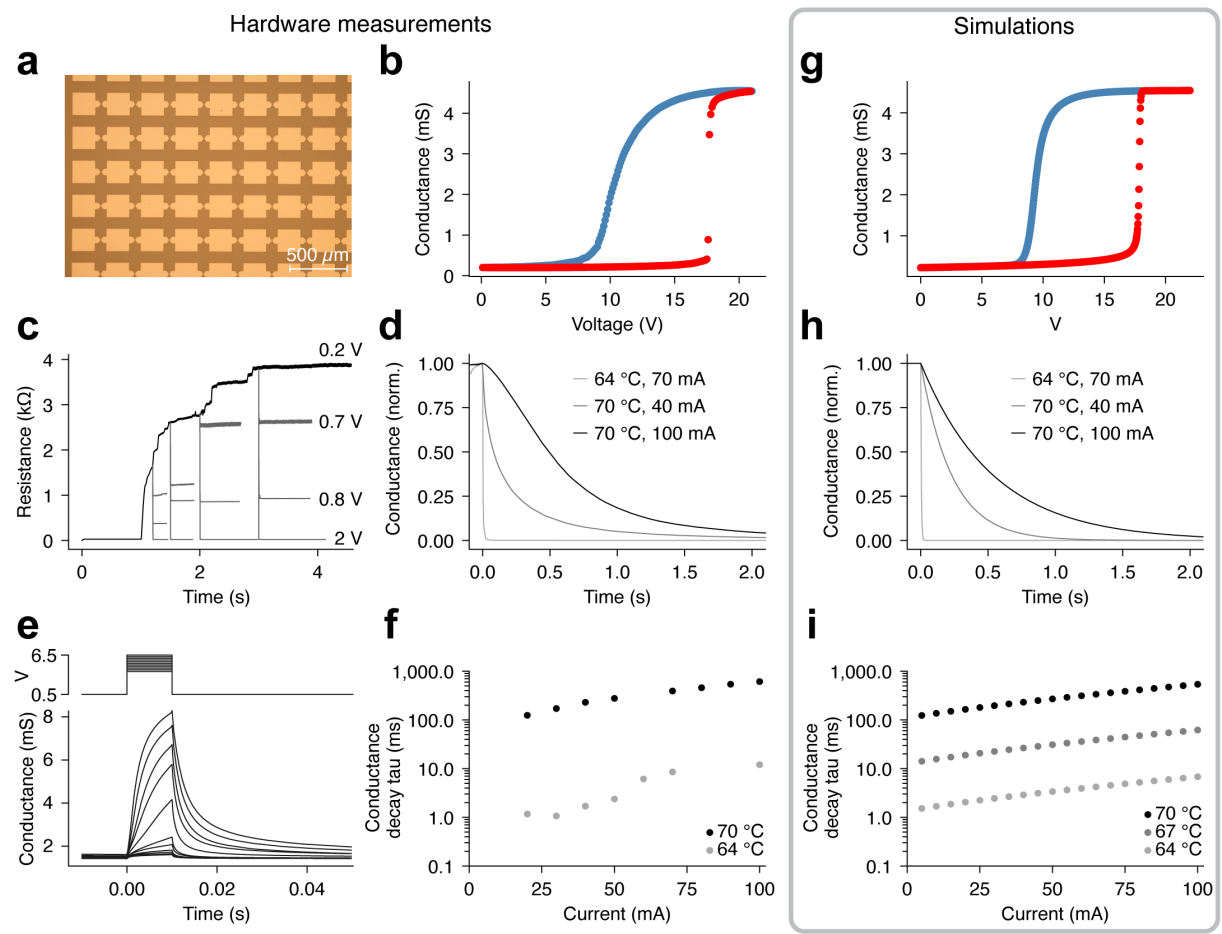

**Fig. 2. Configurable relaxation dynamics and variable resistance of VO$_2$-based devices.**

**a**, Optical image of a VO$_2$ chip, with a grid of gold contacts. Electrical stimulation and recording to control and measure material resistance is performed by placing tungsten electrodes on two neighboring contacts. **b**, The conductance of a VO$_2$ device (70 °C) was measured while an applied voltage was slowly ramped up from 0 to 20 V (red), and then ramped down (blue) over 150 s. Material hysteresis results in different conductance trajectories during activation and relaxation. **c**, A low resistance (high conductance) state of a VO$_2$ device (65 °C) was first activated with a 2 V applied voltage for 1 s. Upon cessation of the applied voltage, the measured resistance of the device increased as the material relaxed towards the non-conducting state (black). In different trials, voltage pulses at a range of amplitudes (0.2 – 2 V) were applied at varying time points during the relaxation time course of the material (grey), which rapidly and stably equilibrated the device to different intermediate resistance states. **d**, Example traces show normalized conductance measurements from a VO$_2$ device. Different combinations of ambient temperature and applied current amplitude resulted in a range of relaxation timescales from 1 ms to > 1 s. **e**, The conductance of a VO$_2$ device (68.5 °C) was measured in response to brief (10 ms) voltage pulses of variable amplitude (4 – 6 V, top), demonstrating that intermediate conductance states can be rapidly accessed by transient stimulation. **f**, Decay time constants from experiments similar to **d**. were estimated by exponential fitting. Increasing either ambient temperature or applied current amplitude increased measured conductance decay time constants. **g-i**, Data is shown from computational model simulations that reproduce physical device properties measured in **b**, **d** and **f**.

which initiates a phase transition within a microdomain of the material, and changes its electrical conductance[34,37]. Scanning thermal microscopy imaging has previously shown that this local temperature change is restricted to the stimulated region of $VO_2$[32,33]. Thus, the conductance states of individual devices in a two-dimensional array can be independently controlled by out-of-plane crossbar electrodes without thermal isolation[43,44,48]. We also verified experimentally that multiple devices on the same $VO_2$ film array can be independently stimulated and measured without influencing the conductance state or relaxation timescale of their neighboring elements (Supplementary Fig. S5e). Recent work has also demonstrated that optimized film deposition methods enable low device-to-device variability as well as stable device characteristics even after millions to trillions of repeated IMT transitions[43,51-53]. These encouraging properties indicate that $VO_2$ is a promising material to consider for integration into hardware to achieve emulation of slow timescale neuronal state variables.

At a given ambient temperature, all neighboring devices on an array share a similar phase relaxation timescale (Supplementary Fig. S5). Thus, in order to emulate multiple different neuronal state variables that decay at a different timescales, a hardware design would have to incorporate multiple $VO_2$ arrays that are thermally isolated from each other and separately thermally regulated. Recent work has demonstrated the feasibility of achieving such local temperature control of CMOS-scale circuit elements[54,55]. However, as an alternative to temperature, it is also possible to configure the properties of devices that utilize phase transitions by selecting other materials and/or through chemical doping [34-36,38,39,45,47,49,56]. In this study, we thus focus primarily on the general configurability of phase relaxation timescales and explore their application to neuromorphic analog computing.

As we showed experimentally, brief pulses of electrical stimulation are sufficient to induce phase transition and increase the conductance of $VO_2$ (Fig. 2e). Thus, our circuit design requires a set of switches to deliver precisely timed "write" voltage pulses (~1 V) to individual elements of a $VO_2$ array (Fig. 3a). Our proposed circuit utilizes separate $VO_2$ device arrays to emulate two different types of neuronal signals – electrical and biochemical (Fig. 1b). When using a $VO_2$ resistor to mimic an ion channel that is a source of electrical current to a neuron, it is not necessary to "read" or record the conductance state of the device. Rather, the voltage across the neuron equivalent circuit is continuously sensed by the $VO_2$ device, and in turn, current generated by the device is continuously integrated by the neuron circuit (Fig. 3a). Importantly, this voltage must be below the threshold voltage for a phase transition so that it does not significantly influence device conductance.

In contrast, for a $VO_2$ resistor to emulate an intracellular biochemical signal that is not a source of electrical current to a neuron, additional circuitry is required to intermittently "read" the conductance state of the device[57]. Our circuit design therefore includes switches to time delivery of "read" voltage pulses to individual devices (Fig. 4a and Supplementary Fig. S5d). When utilizing $VO_2$ devices to mimic biochemical signals involved in synaptic plasticity, it is not necessary to continuously measure the states of all biochemical traces. Rather, only a sparse subset of traces needs to be intermittently monitored whenever a plasticity event is triggered in a postsynaptic neuron (Fig. 4c and Supplementary Fig. S5e).

**Fig. 3.**

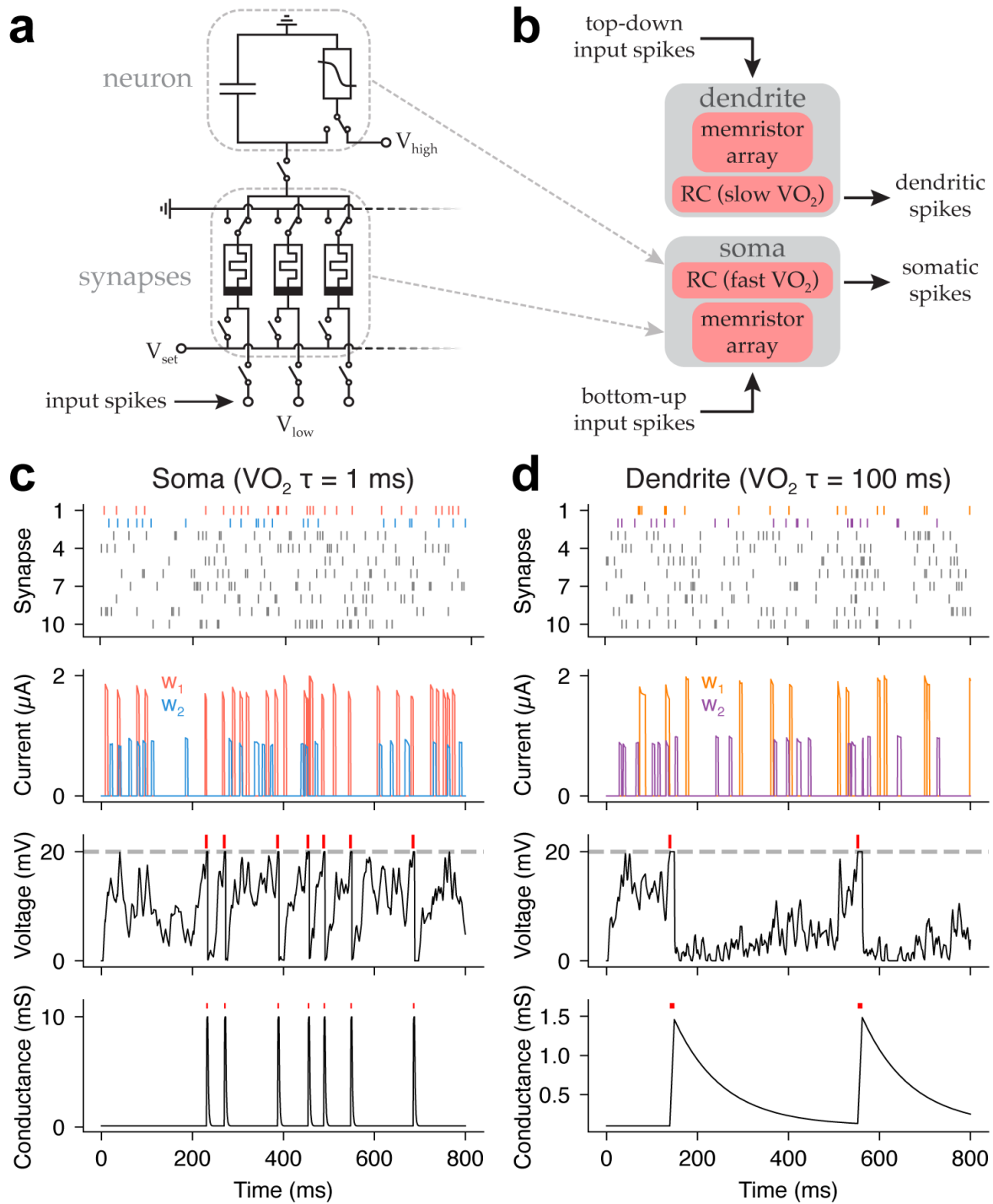

**Fig. 3. Emulating fast somatic spikes and slow dendritic spikes using VO$_2$ phase relaxation.**
**a**, Diagram depicts electrical circuit designed to emulate the membrane voltage dynamics of a neuron receiving synaptic input and generating spike output. This circuit was computationally simulated in **c** and **d**. A VO$_2$ device in its insulating state produces a nonzero membrane conductance reflecting the baseline activity of "leaky" ion channels that drive the neuron's membrane potential towards a resting value (0 mV). A capacitor slows the voltage response of the model neuron in response to time-varying input currents. A presynaptic spike in an input neuron briefly opens a gate to apply a brief low voltage pulse (5 ms, $V_{low}$=40 mV) to an element in a memristor array. The conductances of the non-volatile memristors represent synaptic weights on inputs and can be modified during on-chip learning by applying brief pulses of higher voltage ($V_{set}$=1 V) (Fig. 4). Another gate can be triggered to apply a brief pulse of higher voltage ($V_{high}$=1 V) to transition the $VO_2$ device into a higher conductance state. This results in discharge of the neuron's voltage to ground, and a recovery that depends on the configurable relaxation timescale of the $VO_2$ device. **b**, Diagram depicts two subcircuits of the form in **a**, one representing a neuronal soma compartment that receives bottom-up sensory inputs and emits fast spikes utilizing a $VO_2$ device configured for fast (~1 ms) relaxation, and one representing a dendrite compartment that receives top-down supervisory inputs and emits slow spikes utilizing a $VO_2$ device configured with slow (~100 ms) relaxation. **c**, Computational simulation of a neuronal soma subcircuit. Ten synaptic inputs are activated by Poisson spike trains with mean rate of 30 Hz. Synaptic currents are integrated by the circuit to emulate leaky integration of neuronal membrane potential. When the membrane potential reaches a spike threshold of 20 mV, the conductance state of the $VO_2$ device is increased by a brief voltage pulse, resulting in reset of the membrane voltage for a refractory period determined by the phase relaxation timescale of the $VO_2$ device. Top row shows input spike times. Second row shows synaptic currents from two example inputs with different synaptic weights. Third row shows soma circuit membrane potential. Bottom row shows $VO_2$ conductance. **d**, Same as **c** for simulation of neuronal dendrite subcircuit with slow $VO_2$ relaxation.

      Finally, to achieve on-chip learning, the hardware instantiation of the synaptic weights within a neuromorphic neural network must be modifiable. Recent work has represented synaptic weights in hardware by using phase-change materials as "memristors," which are non-volatile variable resistors with multiple conductance states that can be stably set by applying transient electrical stimulation, and then read with a lower amplitude stimulus[56,58-65]. Storing information in the resistance states of synaptic memory elements, which directly contribute current to the neuronal membrane potential integrator circuit (Fig. 1b and 3a-d), alleviates the need to separately store, retrieve, and transport information to and from a separate on-device memory, saving energy and circuit area[60]. Therefore, our proposed circuit design includes an array of memristors as well as associated crossbars and switches to enable delivery of "read" and "write" voltage pulses to individual synapse elements[66,67].

**Emulating synaptic integration and spiking in separate soma and dendrite compartments**

During biological neuronal signaling, first, a spike in a presynaptic neuron triggers a brief pulse of neurotransmitter release into the extracellular space between neurons, then neurotransmitter-gated ion channels open in the postsynaptic neuron, allowing current to flow across the membrane of the postsynaptic cell. This current flow results in a change in the voltage across the membrane, which is low pass filtered with a cellular time constant that is determined by the resistance and capacitance of the membrane (Fig. 3a). When the voltage in the cell body and axon of a neuron exceeds a threshold, voltage-gated channels open, leading to an all-or-none spike. Neurons also contain tree-like dendrite compartments where most excitatory synapses are received. Dendrites also express voltage-gated channels, including calcium channels that can generate long duration events called dendritic calcium spikes, or plateau potentials[40-42].

Given the key role that dendritic spikes play in BTSP, we next sought to apply our above-described neuromorphic circuit design to emulate the integration and spiking behavior of separate somatic and dendritic compartments. We computationally simulated the circuit as follows: each compartment included a capacitor to mimic the leaky and integrative properties of a neuronal membrane, and a $VO_2$-based volatile resistor to mimic the ion channels that reset the membrane potential after a spike and regulate the duration of the refractory period following each spike[43-45,48] (Fig. 3a-d). Both the soma and dendrite compartments received synaptic inputs that were driven by presynaptic spikes, with variable weights that were determined by the resistances of an array of modifiable non-volatile memristors. The somatic compartment integrated bottom-up input carrying information about a sensory stimulus, and was connected to a $VO_2$ device configured for fast (~1 ms) relaxation to implement a short spike duration and brief refractory period (Fig. 3a-c). The dendritic compartment integrated top-down input carrying information that was supervisory for learning, and was connected to a $VO_2$ device configured for slow (~100 ms) relaxation to implement a broader calcium spike and a longer refractory period (Fig. 3a,b,d).

Each time a presynaptic spike arrived, a synaptic current was generated by opening a gate and applying a brief low voltage "read" pulse to the memristor associated with that synaptic input. Currents from the array of synaptic inputs were then integrated by the neuronal compartment RC circuit, with the resistance determined by a $VO_2$ device in its insulating state. When integration of synaptic inputs caused the voltage of the neuronal compartment to increase beyond a predetermined spike threshold, a gate was opened to deliver a brief activating voltage pulse to the $VO_2$ device, increasing its conductance. This resulted in a rapid discharging of the neuron compartment capacitor through the high $VO_2$ conductance, and a decrease in the model membrane voltage. Then, as the $VO_2$ device relaxed back to its insulating state, currents arriving from incoming synaptic inputs leaked to ground and prevented the voltage across the neuron compartment from reaching spike threshold again for a refractory period set by the relaxation time constant of the $VO_2$ device. As mentioned, in our simulations we set the relaxation timescale of the $VO_2$ device in the soma compartment to be fast (~1 ms) to achieve a short refractory period (Fig. 3c), and set it to be slow (~100 ms) in the dendrite compartment to achieve a longer refractory period (Fig. 3d). These results demonstrate that a hybrid electrical circuit incorporating phase transition materials with engineered relaxation timescales can be used to emulate multiple timescales of biological electrical signaling in neuromorphic units with separate soma and dendrite compartments.

**Temporal credit assignment for one-shot learning**

Next, we built upon our neuromorphic circuit design and computationally simulated using additional arrays of $VO_2$ devices to emulate the ultraslow (~1 s) relaxation times of intracellular biochemical signals related to synaptic plasticity (Fig. 4a,c and Supplementary Fig. S6b; see Methods). In this extended circuit, each time a spike was generated in a presynaptic neuron, in addition to triggering synaptic currents in postsynaptic neurons (Fig. 3a,c,d), it also incremented an analog synaptic eligibility trace (ET) by triggering a gate to apply a brief voltage pulse to an element in another array of $VO_2$ devices (ET array) (Fig. 4a). The exponential decay of each ET was determined by the relaxation timescale of the $VO_2$ array, which was set to ~1.5 s in order to achieve the long timescale of associative learning characteristic of the BTSP learning rule (Fig. 4a,c and Supplementary Fig. S6a,b). In addition, whenever a calcium spike was emitted by a neuronal dendrite compartment, an analog instructive signal (IS) was incremented by opening a gate to deliver a brief voltage pulse to an element in the fourth array of $VO_2$ devices (IS array) (Fig. 4a). The relaxation timescale of the IS array was set to ~500 ms, which enables synapses activated within an extended time window after a calcium spike to also undergo plasticity[24] (Fig. 4a,c and Supplementary Fig. S6a,b). For the duration of the IS in a postsynaptic neuron, the ET traces at all incoming synapses onto that cell were intermittently measured by applying "read" pulses to a subset of the ET array (Fig. 4c). Finally, synaptic weight updates were calculated as a function of the temporal overlap between the IS and each ET[24] (Supplementary Fig. S6; see Methods), and "write" pulses were delivered to the corresponding synaptic memristors to modify their conductances and store the associative memory (Fig. 4c).

We next used the above-described scheme to simulate a biological experiment in which the membrane voltage of a hippocampal neuron was recorded while a mouse ran on a treadmill decorated with distinguishing sensory cues[24]. We modeled a population of presynaptic neurons with spatially-tuned activity such that each neuron generated spikes at a different subset of locations along the track (Fig. 4b). These inputs were integrated in the soma compartment of a model postsynaptic neuron (Fig. 4g). Initially the synaptic weights were set to a low value, resulting in a postsynaptic membrane potential that did not exhibit any location selectivity or generate any spikes (Fig. 4g). During one single trial, the dendrite compartment received a supervisory signal at a particular location along the track (Fig. 4c), causing the neuron to emit a dendritic calcium spike and induce BTSP. In Figure 4c, the ET associated with an example synaptic input is shown alongside the IS associated with the postsynaptic neuron. These time-varying biochemical traces were represented by the intrinsically decaying conductances of corresponding ET and IS $VO_2$ devices. The timing of "read" and "write" pulses delivered to these devices are marked (Fig. 4c). After the IS completed its relaxation, weight updates were calculated for all synapses according to the BTSP learning rule (Supplementary Fig. S6a,c; see Methods), and their weights were updated by "write" pulses delivered to their associated memristors (Fig. 4c,f). On the next trial, as the spatially-selective inputs were activated, the membrane potential in the soma compartment of the postsynaptic neuron slowly ramped up, and began to emit somatic spikes at a high rate at spatial locations on the track nearby the location where the dendritic calcium spike had been emitted in the previous trial (Fig. 4g). This rapid modification to the synaptic weights and resulting somatic membrane potential after a single trial recapitulates the rapid change in membrane potential previously observed in the biological data[24] (Fig. 4d,e). These results demonstrate that relaxation dynamics in phase transition materials can be used to achieve temporal credit assignment for implementation of one-shot learning on neuromorphic hardware that generates temporally extended stimulus representations in a single trial.

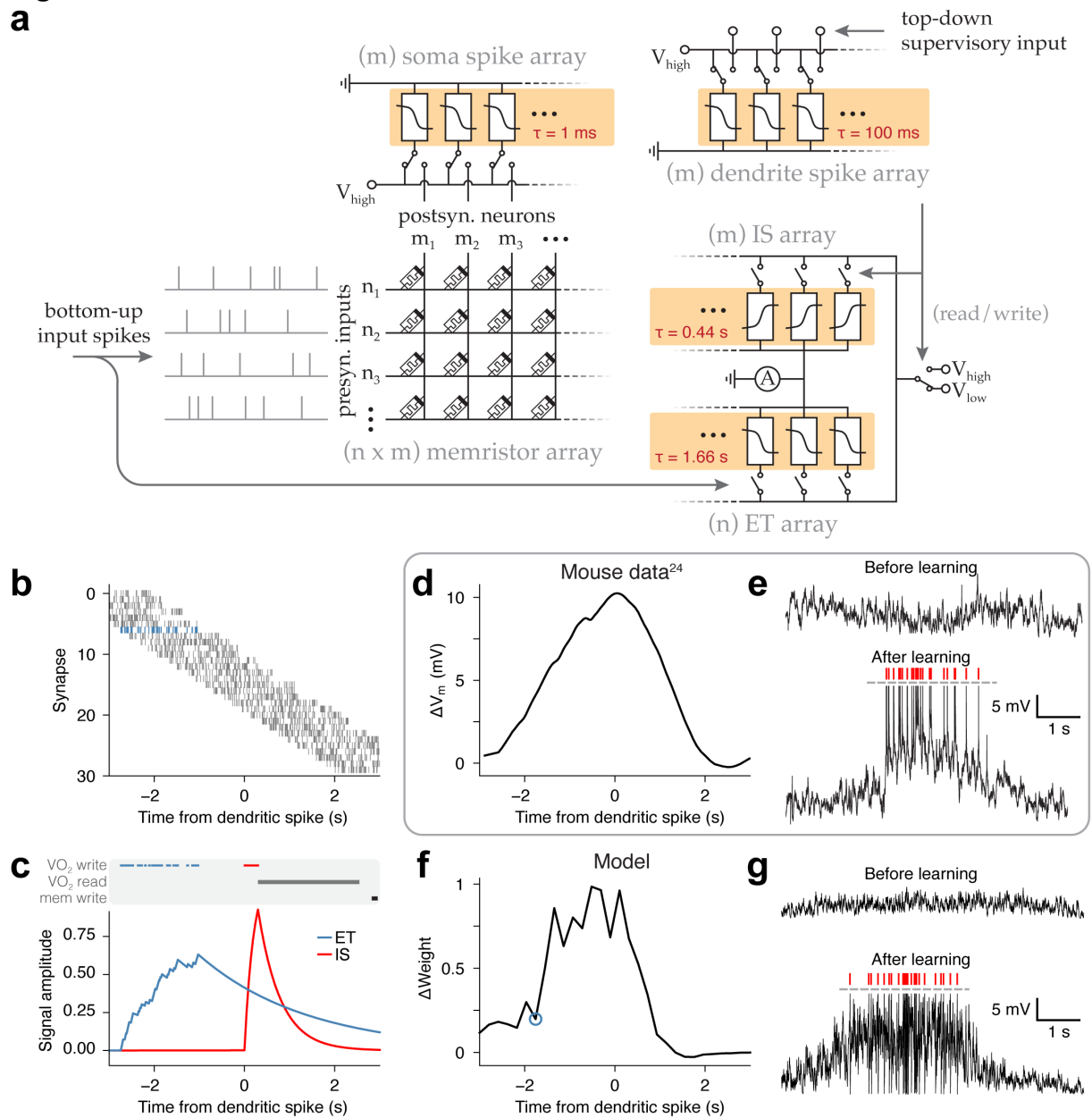

**Fig. 4. Phase relaxations enable temporal credit assignment for neuromorphic one-shot learning.**
**a**, Diagram illustrates a neuromorphic circuit implementing a neuronal network containing a layer of *m* postsynaptic neurons receiving synaptic input. The circuit incorporates four separate VO$_2$ arrays, each configured to relax at a different timescale to emulate a different biological feature required for behavioral timescale synaptic plasticity (BTSP). As in Fig. 3, incoming presynaptic spikes from *n* presynaptic neurons result in sparse activation of an *n* x *m* memristor array. Bottom-up inputs are integrated by soma subcircuit emitting fast spikes with a fast VO$_2$ array, and top-down inputs are integrated by a dendrite subcircuit emitting slow spikes with a slow VO$_2$ array. Presynaptic spikes also activate elements in an ultraslow VO$_2$ array that emulates a biochemical eligibility trace (ET) which increments with each successive spike and marks recently activated synapses as eligible for plasticity. Postsynaptic dendritic spikes activate elements in an ultraslow VO$_2$ array that emulates a biochemical instructive signal (IS) which changes the strengths of synapses with nonzero ETs. **b-g**, The circuit in **a** was computationally simulated to emulate a biological experiment in which the intracellular membrane potential of a neuron in the rodent hippocampus was recorded during spatial navigation. A dendritic spike emitted in the neuron induced BTSP, which changed synaptic strengths to store a memory of spatial position. This was reflected in a spatially-selective increase in membrane potential and spiking after learning (**d**, **e**). Experimental data is reproduced from Milstein et al., 2021[24]. **b**, The Poisson spike times of presynaptic neurons during an example lap of simulated running on a linear track are shown. Each unit's activity is spatially and temporally restricted to a 2 s period with a mean rate of 30 Hz during a 6 s lap. **c**, During simulated lap running, each time a presynaptic spike arrived, the conductance of an element in the ET $VO_2$ array was incremented by a voltage pulse (timing of $VO_2$ write pulses indicated by blue ticks). The resulting normalized conductance of an example ET is shown in blue for the input with spike train colored blue in **b**. A supervisory input to the postsynaptic neuron's dendrite caused it to emit a dendritic calcium spike in the center of the track. A device in the IS $VO_2$ array was activated by a voltage pulse (timing of write pulse indicted by red tick). The resulting normalized conductance of the IS is shown in red. While the IS is nonzero, its value and the values of ET at all incoming synapses are read with intermittent voltage pulses (timing of $VO_2$ read pulses indicated by gray ticks). These values are used to compute weight updates according to the BTSP learning rule (Methods). **d**, **e**, Experimentally observed spatially-restricted increase in neuronal membrane potential (**d**,**e**) and spiking (**e**) above baseline from before to after one-shot learning induced by a dendritic calcium spike[24]. **f**, **g**, Results of computational simulation of BTSP using neuromorphic hardware for comparison to **d**, **e**. **f**, The change in synaptic weight from before to after a dendritic spike is plotted for inputs activated at a range of time delays to the dendritic spike. **g**, Spatially-restricted increase in simulated somatic membrane potential and spiking above baseline from before to after one-shot learning induced by a dendritic calcium spike.

**Application to real-time spatial navigation and reinforcement learning**
The above results demonstrated that a metal-insulator-transition material exhibiting slow relaxations can be used to mimic intracellular signaling important for a form of synaptic plasticity that associates events across long time delays. We next considered whether a network of neuromorphic units with this capability could be used to implement a new sample-efficient reinforcement learning algorithm. We simulated a virtual agent searching for a reward in a spatial environment and used the population activity of a network of neuromorphic units to guide the movements of the agent (Methods). At each spatial position, a different presynaptic input to the network was activated, and simulated $VO_2$ variable resistors were used to track a slowly decaying ET at each input. Then, at each spatial position a different postsynaptic neuron in the network was triggered to evoke a dendritic calcium spike, and another set of simulated $VO_2$ devices were used to track a slowly decaying IS in each postsynaptic neuron (Figs. 1c and 4a,X). These signals were used to compute synaptic weight updates at each pair of connected pre- and postsynaptic neurons according to the BTSP learning rule (Supplementary Fig. S6; see Methods), which stored a spatial map of the environment in the network.

      During unidirectional running on a linear treadmill, the BTSP learning rule generates spatial receptive fields in hippocampal neurons that have an asymmetric shape - they extend backwards in time from their peak for multiple seconds, but extend forward in time for a shorter duration[23,24] (Fig. 4d-g and Supplementary Fig. S6c). Representations with this shape are referred to as "predictive representations," as the activity of a neuron begins to ramp up before the animal reaches the neuron's preferred location, and can be used to decode the future position of the animal[68]. In the reinforcement learning field, when a population of neurons expresses predictive representations with this temporally skewed shape, the vector of activity of the entire population is referred to as a "successor representation"[69-72]. These types of predictive spatial representations are particularly well-suited for goal-directed learning, because the problem of associating spatial positions with rewards can be divided into two independent processes - one neural circuit can learn the topology of the environment and store information about the temporal sequences of environmental features that an animal or virtual agent tends to experience in an order, and another process can separately record the locations of the encountered rewards. Then a navigation algorithm can consult both the topology of the environment and the value of each position to construct an efficient trajectory towards a reward (Fig. 1c). This is particularly advantageous under conditions of

changing reward value, as the reward representation can be updated independently of the map of space.

Typically, successor representations are learned with what is known as the "temporal difference" learning rule (TD)[70-73]. This algorithm requires multiple iterations of exploring an environment to construct a successor representation, and the shape of the representation is characteristically skewed by the order that positions in the environment are visited. If it is the policy of an animal or agent to run at a constant velocity in a single direction, then the successor representation learned by TD will skew in the direction of travel, predicting the next step of the animal[69]. Whether the brain implements a version of the TD learning rule, or utilizes other plasticity mechanisms during spatial map formation is an active area of research[74-76]. However, the fact that the BTSP learning rule generates temporally-extended predictive receptive fields in as few as one single trial of experience suggests that it could represent a new biologically-plausible mechanism for goal-directed learning. Here we demonstrate that when the action policy of a virtual agent is fixed to run in a single direction along a stereotyped linear trajectory (Fig. 5a), the BTSP learning rule causes a network of neurons to rapidly learn a temporally asymmetric successor representation in 4-fold fewer trials than the TD learning rule (Fig. 5b,c). Note that this difference does not reflect a difference in learning rates; regardless of learning rate, the TD learning rule can only update successor representations one state at a time, and so always requires multiple trials to construct a temporally-extended representation. To verify that the long timescale of the BTSP learning rule, enabled by the long timescale relaxation dynamics of the $VO_2$ devices, is important for these results, we also compared to a standard short timescale Hebbian learning rule, which failed to learn a successor representation even after many repeated trials (Fig. 5b,c).

We then extended the test of our simulated neuromorphic network on a goal-directed reinforcement learning task with a single reward located at a fixed position within a two-dimensional environment. A virtual agent was placed at a fixed start position within the environment, and different learning algorithms were used to determine the direction of the agent's movements. Upon encountering the fixed reward location, the agent was reset to the start position, and this was repeated for multiple trials. On the first trial, all spatial positions were assigned equal value, and the action policy of the agent was set to move to any adjacent position with equal probability, avoiding walls and barriers (Fig. 5d, left, red trajectories). During exploration, depending on the synaptic plasticity rule, neurons within the network changed their spatial activity (Fig. 5d, center and right). Once the agent encountered the rewarded position, the action policy began to consult the spatial activity of the neurons in the network that are active near the reward to determine the direction of travel. This caused the agent to become biased towards repeating trajectories that have previously led to the goal (Fig. 5d, left, blue trajectories; see Methods). When the network was trained with the BTSP learning rule, it produced temporally-extended predictive representations in a single trial (Fig. 5d, center), and resulted in an up to 4-fold reduction in the number of trials to learn an efficient path to the reward location compared to the standard TD and Hebbian learning rules (Fig. 5e-h). We note that while in this simulation, plasticity was triggered in a single layer of neurons by explicit teaching signals, previous work has shown that dendritic gating of plasticity extends effectively to multilayer networks with top-down feedback to neuronal dendrite compartments[9,24,77,78]. In summary, these results demonstrate that long timescale analog signaling using phase transition materials can enable neuromorphic hardware to implement powerful biology-inspired learning rules that build predictive representations with benefits for reinforcement learning.

**Fig. 5.**

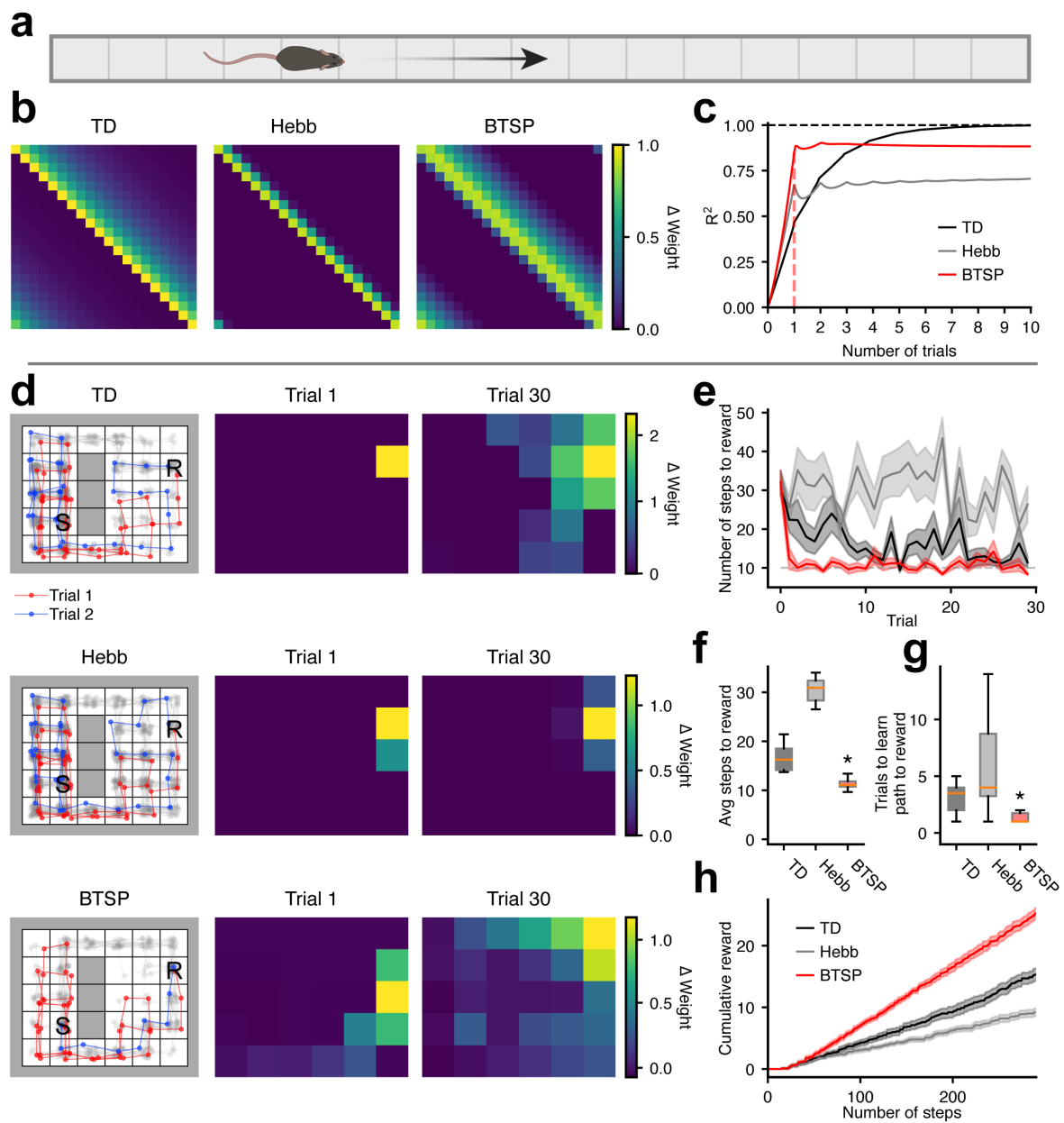

**Fig. 5. Phase relaxations enable temporal credit assignment for spatial navigation reinforcement learning in a neuromorphic neural network.**
**a**, The neuromorphic network schematized in Fig. 1c and Fig. 4a was simulated using the plasticity model shown in Fig. 4b-g. A virtual agent was simulated as moving in a single direction at constant velocity along a linear track. **b**, Simulations were conducted using different learning rules to adjust the synaptic weights within the network during repeated trials of the agent navigating the track. After 10 repeated trials, the standard temporal difference (TD) learning rule stored in the synaptic weights of the network a predictive "successor representation" of the sequence of spatial positions that the agent experienced (left). While the BTSP learning rule also produced a temporally extended, asymmetric representation (right) similar to the TD rule, a standard Hebbian learning rule did not (center). **c**, At each trial during learning, the distributions of synaptic weights within the network were compared to the final successor representation produced by the TD learning rule ($R^2$ was calculated from Pearson's correlation between the weights on a given trial and the weights produced by the TD rule after trial 10). While the TD learning rule (black) takes 4 trials to construct the successor representation, the BTSP learning rule (red) produces its successor-like representation in a single trial. The Hebbian rule does not learn a successor representation even after 10 trials. **d**, The above model was applied to simulated navigation in a two-dimensional environment containing a reward at a fixed position, and the performance of the above three learning rules were compared. Left: the trajectories of the virtual agent across 30 trials are depicted (grey), with the first (red) and second (blue) trials highlighted. Center: the synaptic weights of spatial inputs onto a single neuron in the network are shown after the first trial of exploration. Right: the weights onto the same neuron are shown after 30 trials. **e**, The number of steps taken to reach the rewarded position is shown for each trial and learning rule. Shading reflects mean ± SEM across 10 independent instances of each network simulation. **f**, The number of steps to reward is averaged across trials and compared across learning rules. Asterisk reflects statistical significance (two-tailed t-tests; BTSP vs. TD: $p<0.0001$, BTSP vs. Hebb: $p<0.0001$; Bonferroni corrected for multiple comparisons). **g**, The number of trials taken to learn an efficient path to reward (<10 steps) is compared across learning rules. Asterisk reflects statistical significance (two-tailed t-tests; BTSP vs. TD: $p<0.0042$, BTSP vs. Hebb: $p<0.0135$; Bonferroni corrected for multiple comparisons). **h**, The cumulative number of rewards obtained is related to the number of steps taken during learning, and compared across learning rules.

**Conclusion**
In this study, we demonstrated that metal-insulator phase transition kinetics can be dynamically controlled and configured to match the heterogeneous timescales of analog neuronal signaling important for behavioral timescale synaptic plasticity. Our simulations indicate that material phase relaxations can be used to perform temporal credit assignment for associative learning across seconds-long time delays in neuromorphic hardware. We showed that a neuronal network exploiting these material properties can implement a powerful biology-inspired reinforcement learning algorithm, improving the sample efficiency of real-time maze learning by up to 400%. These findings highlight the versatility and potential utility of phase transition materials at the interface between neuroscience, electrical engineering, and machine learning.

**Methods**

*Fabrication of VO$_2$ devices*
VO$_2$ thin films of 100-nm thickness were deposited on c-plane Al$_2$O$_3$ substrates at 650 °C by using a radio-frequency (RF) magnetron sputtering approach (AJA International). A V$_2$O$_5$ ceramic plate which had a purity of 99.9% was used as the sputtering target. The RF sputtering power was set at 100 W for a growth rate of 0.6 nm/min. A chamber pressure of 5 mTorr with an Ar/O$_2$ mixture ratio of 99.9/0.1 was used. The in-plane VO$_2$ devices were fabricated by photolithography with a photo resist of AZ 1518 as a mask layer. The pattern for electrodes was directly written by Heidelberg MLA150 Maskless Aligner. The electrode layer, 100-nm-thick Pt, was deposited by e-beam evaporation and then lifted off by PG-Remover at 80 °C. The gap between two Pt electrodes was 100 $\mu$m.

*Electrical measurements and structural characterization*
The electrical measurements were performed on a micromanipulator probe station with temperature range from RT to 200 °C. The electric properties of the VO$_2$ device were measured by a Keithley 2461 SourceMeter. The XRD profile was obtained with the Panalytical Xpert X-Ray Diffractometer at room temperature by θ-2θ scans. The optical images were obtained with Zeiss Axioscope 40 Optical Microscope.

*Hardware experimental measurements from an array of VO$_2$ devices*
We verified with hardware measurements that neighboring VO$_2$ devices on an array could be subject to intermittent "read" and "write" voltage pulses without influencing each other's conductance or relaxation timescales (Supplementary Fig. S5). Platform temperature was set to 70 °C to achieve a VO$_2$ conductance relaxation timescale >100 ms. We employed a time-division technique to stimulate and measure the time-varying conductance of three distinct VO$_2$ devices. Given a constant DC voltage source, the conductance of a single device can be calculated from the current measured by an ammeter, provided that only a single device is being recorded at a time. We connected the three devices to two DC voltage sources, one providing low voltage for conductance measurements, and one providing higher voltage to induce the insulator-metal phase transition. A set of six switches determined which device was in either the stimulation or recording mode at a given time. An Arduino board and a high-current Darlington transistor array toggled the switch states every 2 ms.

*Computational simulation of VO$_2$ device properties*
VO$_2$ devices were modelled phenomenologically as volatile time-varying variable resistors. Given an insulating state resistance $R_{ins}$ and a metal state resistance $R_{metal}$, an applied control current *I* determined a target equilibrium resistance $R_{eq}$ (Ω) according to a relationship that was fit to data from hardware measurements (Fig. 2b,g):

(1) $\quad R_{eq} = (R_{ins} - R_{metal}) \cdot e^{(-11 \cdot I)} + R_{metal}$

The equilibrium conductance $g_{eq}$ (S) is the reciprocal of $R_{eq}$:

(2) $\quad g_{eq} = 1/R_{eq}$

We assumed that the current required to induce insulator-metal-transition is <1 mA for devices with gap sizes <2.5 μm (Supplementary Fig. S3 and S4).

Given an operating temperature *T*, if the equilibrium conductance $g_{eq}$ was greater than the instantaneous conductance *g*, the model device conductance approached $g_{eq}$ with an exponential rise time constant $\tau_{rise}(T, I)$ (ms) that was fit to data from hardware measurements (Supplementary Fig. S3b,c). In the absence of applied current, or when $g_{eq}$ was less than *g*, the conductance approached $g_{eq}$ with an exponential decay time constant $\tau_{decay}(T, I)$ (ms) that was fit to data from hardware measurements (Fig. 2d,h,f,i and Supplementary Fig. S3b,c,f,g):

(3) $\quad \tau_{rise}(I, T) = (-60 \cdot I + 175) \cdot \alpha(T)$

(4) $\quad \tau_{decay}(I, T) = (227.7 \cdot e^{\frac{I}{0.87}} - 116.7) \cdot \alpha(T)$

(5) $\quad \alpha(T) = 20/(1 + e^{\frac{74-T}{1.359}})$

During simulations, the conductances of VO$_2$ devices were evolved with a timestep of dt=1 ms.

*Computational simulation of neuronal input summation and spiking*

The membrane potential *V* of a neuron compartment receiving synaptic inputs was modeled as an equivalent electrical circuit (Fig. 3a). The time course of change in membrane potential $\frac{dV}{dt}$ was slowed by a capacitor *C*=1 μF in series with a VO$_2$ variable resistor with insulating resistance $R_{ins}$=10 kΩ and metal resistance $R_{metal}$=100 Ω. Upon arrival of presynaptic spikes, synaptic currents $I_{syn}$ were generated by applying 5 ms voltage pulses (40 mV) to elements in a memristor array. The rate of change in membrane potential $\frac{dV}{dt}$ was defined as follows:

(6) $\quad C\frac{dV}{dt} = -V/R_{VO_2} + I_{syn}$

In neuronal soma compartments, VO$_2$ devices in a soma spike array were simulated at a temperature of 62.0 °C, producing conductances that decayed with a time constant of ~1 ms. When V in a soma compartment exceeded a spike threshold of 20 mV, a device in the soma spike array was activated with an applied current of 1 mA for 3 ms. In neuronal dendrite compartments, VO$_2$ devices in a dendrite spike array were simulated at a temperature of 69.3 °C, producing conductances that decayed with a time constant of ~100 ms. When V in a dendrite compartment exceeded a spike threshold of 20 mV, a device in the dendrite spike array was activated with an applied current of 1 mA for 10 ms. Simulations were evolved with a timestep of dt=0.1 ms (Fig. 3b,c).

*Computational simulation of BTSP*
The above computational model of $VO_2$ device properties was used to implement variable resistors, and the time-varying conductances of the modeled devices were interpreted as biochemical signals required for synaptic plasticity. In Fig. 4b,c,f,g, a set of 30 presynaptic inputs were modeled as spike times generated by a Poisson process with a mean rate of 30 Hz. Each input was active for a 2 s period during a 6 s simulation trial (Fig. 4b). An array of $VO_2$ devices was used to represent biochemical eligibility traces (ET) that mark synapses as eligible for synaptic plasticity. This ET array was simulated at a temperature of 74.3 °C, producing conductances that decayed with a time constant of ~1.66 s. Upon arrival of each presynaptic spike, a device in the ET array was activated with an applied current of 1 mA for 20 ms. In Fig. 5, a set of 20 (Fig. 5a) or 30 (Fig. 5b) presynaptic inputs were modeled as binary increases in firing rate from 0 to 1 for a duration of 400 ms, and the corresponding ET $VO_2$ devices were activated with an applied current of 1 mA for 400 ms.

Another array of $VO_2$ devices was used to represent biochemical instructive signals (IS) that gated plasticity in postsynaptic cells. This IS array was simulated at a temperature of 70.8 °C, producing conductances that decayed with a time constant of ~0.44 s. Upon generation of a dendritic calcium spike in a postsynaptic neuron, a device in the IS array was activated with an applied current of 1 mA for 300 ms. Conductances measured from the ET and IS $VO_2$ arrays were offset and normalized to produce values between 0 and 1. Whenever an IS was nonzero, it converted any nonzero synaptic ETs into changes in synaptic strength according to a previously reported mathematical model[24], which we briefly summarize below.

For an $ET_i$ at presynaptic neuron *i*, and an $IS_j$ at postsynaptic neuron *j*, the change in synaptic weight $W_{ji}$ evolved in time as follows:

(8) $\quad \frac{dW_{ji}}{dt} = \lambda \cdot [(W_{max} - W_{ji}) \cdot k^+ \cdot q^+(ET_i \cdot IS_j) - W_{ji} \cdot k^- \cdot q^-(ET_i \cdot IS_j)]$

Where learning rate $\lambda$=1.2, $W_{max}$=4.68, $k^+$=1.1097, and $k^-$=0.425. As previously reported[24], $q^+$ and $q^-$ were sigmoidal functions, stretched and scaled to intercept the coordinates (0, 0) and (1, 1). $q^+$ had a slope of 4.405 and a threshold of 0.415, and $q^-$ had a slope of 20.0 and a threshold of 0.026. Weight updates were computed every 10 ms during simulation and aggregated, resulting in a single weight update at the end of the one-shot learning trial.

*Computational simulation of spatial navigation reinforcement learning*
In Fig. 5, the movements of a virtual agent navigating a spatial environment were simulated. We compared three approaches to spatial navigation and goal-directed learning – the standard temporal difference algorithm (TD)[70-73], a short timescale Hebbian learning rule, and the above-described neuromorphic hardware-compatible implementation of the BTSP learning rule. A one-dimensional linear track (Fig. 5a-c), and a two-dimensional square arena (Fig. 5d-h) were divided into discrete spatial positions, or "states" *s* that were occupied for 400 ms as the agent moved at a constant velocity. For an environment with *n* total number of states, the TD rule updates an *n* x *n* matrix *M* called the "successor matrix" that stores information about the temporal proximity between states. When the agent is in a state $s_i$, a row *i* of the successor matrix is a vector of length *n*, and each column *j* contains a learned value that represents the likelihood that the agent will visit state $s_j$ at some point in the future. During occupancy of state $s_i$, values in row *i* and column *j* in the matrix *M* were updated as follows:

(9) $\quad M_{i,j} = \gamma \cdot M_{i+1,j} + \begin{cases} 0, & i \neq j \\ 1, & i = j \end{cases}$

where the temporal discount factor $\gamma = 0.75$.

For the Hebbian learning rule, an analog of the successor matrix *M* was stored in the synaptic weights between two populations of neurons. Each unit *i* in a population of *n* presynaptic neurons was assigned to represent a state $s_i$. The firing rate $r_i^{pre}$ of unit *i* was set equal to 1 when the agent occupied state $s_i$, and set to 0 in all other states. Each unit *j* in a population of *n* postsynaptic neurons received synaptic connections from all *n* presynaptic neurons with weights $W_{ji}$ from presynaptic unit *i* to postsynaptic unit *j*. The firing rate $r_j^{post}$ of each postsynaptic unit *j* depended on a weighted sum of its inputs. In addition, each postsynaptic unit *j* was assigned a state $s_j$ where the firing rate $r_j^{post}$ was forced to increase by a value of 1. When the agent occupied state $s_i$, the firing rate $r_j^{post}(s_i)$ of each postsynaptic unit *j* was computed as follows:

(10) $\quad r_j^{post}(s_i) = f(\sum_i^n (W_{ji} \cdot r_i^{pre}(s_i))) + \begin{cases} 0, & i \neq j \\ 1, & i = j \end{cases}$

where f was a rectified sigmoidal activation function:

(11) $\quad f(x) = \begin{cases} 0, & x < 0 \\ \frac{2}{1+e^{-x}} - 1, & x \geq 0 \end{cases}$

During simulated navigation, the synaptic weights $W_{ji}$ were updated according to the following rule similar to spike-timing-dependent plasticity (STDP)[79]: presynaptic activity in the state preceding postsynaptic activity, or presynaptic activity in the same state as postsynaptic activity increased synaptic weights, while presynaptic activity in the state following postsynaptic activity decreased synaptic weights. The change in synaptic weight $\Delta W_{ji}$ at the connection from presynaptic unit *i* to postsynaptic unit *j* when the agent occupied state $s_k$ was computed as follows:

(12) $\quad \Delta W_{ji} = \lambda \cdot [0.5 \cdot r_i^{pre}(s_{k-1}) \cdot r_j^{post}(s_k) + r_i^{pre}(s_k) \cdot r_j^{post}(s_k) - 0.5 \cdot r_i^{pre}(s_k) \cdot r_j^{post}(s_{k-1})]$

where the learning rate $\lambda$=0.04.

For the BTSP learning rule, rather than forcing each postsynaptic neuron *j* to increase its firing rate in state $s_j$, each unit was forced to evoke a dendritic calcium spike in its assigned state $s_j$. As described above, presynaptic activity incremented an ET at each synapse, and dendritic spikes incremented an IS in each postsynaptic neuron. ET and IS traces were emulated using the model of $VO_2$ devices as before, and updated every dt=1 ms. Weights $W_{ji}$ were initialized to a baseline value of 1. Synaptic weight updates were computed every 10 ms and aggregated, resulting in one weight update per state (with 400 ms occupancy).

During navigation on the one-dimensional track, the action policy or behavior of the agent was set to move in a single direction. However, on the two-dimensional track, the agent could choose between 4 possible directions of movement, avoiding walls and barriers, and the choice of movement was biased by the past experience of the agent and the location of encountered rewards. To keep track of expected reward, a vector $R(s_i)$ of length number of states $n$ was updated in each state $s_i$ depending on the presence or absence of reward. For the TD algorithm the expected value $V(s_i)$ of each state $s_i$ was computed as follows:

(13) $\quad V(s_i) = \sum_{k}^{n}(M_{i,k} \cdot R(s_k))$

Before encountering any reward, all states had equal value of zero, so all possible actions were selected with equal probability. In our simulations, only a single reward was placed at a single fixed location in state $s_R$. After encountering the rewarded position at least once, the reward value $R(s_R)$ of state $s_R$ was set equal to 1, and the expected value $V(s_i)$ of each state reduced to:

(14) $\quad V(s_i) = M_{i,R}$

For the Hebb and BTSP algorithms, the value was calculated analogously from the synaptic weights $W$ within the neural network:

(15) $\quad V(s_i) = \sum_{k}^{n}(W_{ki} \cdot R(s_k))$

which for the special case of a single fixed reward reduced to:

(16) $\quad V(s_i) = W_{Ri}$

where $W_{Ri}$ is the synaptic weight of the connection from the presynaptic unit assigned to represent state $s_i$ onto the postsynaptic unit assigned to represent the rewarded position $s_R$. Then, to select the next spatial position, an "epsilon greedy" policy[70-73] consulted the expected value of all states adjacent to the currently occupied state, and selected the state $s_k$ with the maximum expected value $V(s_k)$ with probability 1-$\varepsilon$, or selected a random adjacent state with probability $\varepsilon$, where $\varepsilon = 0.2$.

## Data Availability

The experimental and model simulation data that support the findings of this study are available from the corresponding authors upon reasonable request. The code necessary to reproduce model simulations will be made publicly available upon publication.

## Acknowledgements

This work was supported by funding from Rutgers Biomedical and Health Sciences (A.D.M.), a fellowship from the European Molecular Biology Organization (A.R.G.), AFOSR grant FA9550-22-1-0344 (C.G. and S.R.), and the Smith Family Foundation (C.G.). RBS measurements were carried out at the Laboratory for Surface Modification (LSM) of Rutgers University.

**Author contributions**
Conceptualization: C.G., S.R., A.D.M.; supervision: C.-T.M.W., S.R., A.D.M.; device fabrication: H.Y., Y.Y.; experimental design: Y.Y., A.R.G., M.Z., S.R., A.D.M.; experiments and data collection: Y.Y., R.S.B., M.Z., A.R.G.; data analysis: Y.Y., A.R.G., R.S.B.; computational modeling: A.R.G., A.D.M.; writing—original draft: A.R.G., A.D.M.; writing—review & editing: S.R., A.D.M.

**Ethics declarations**
Competing interests

The authors declare no competing interests.

**Supplementary Information**

**Structural and electrical characterization of VO$_2$ devices**
To analyze the structure of VO$_2$ thin films, X-ray diffraction (XRD) $\theta$-$2\theta$ scans were performed on VO$_2$/Al$_2$O$_3$ samples with and without Pt. As shown in Supplementary Fig. S1a, only the (002) peak of the VO$_2$ thin film with monoclinic phase[1] and (006) peak of Al$_2$O$_3$ substrate appear, which indicates that the VO$_2$ is oriented along the out-of-plane direction of Al$_2$O$_3$. The epitaxial orientation is consistent with previous reports[2] that the deposited VO$_2$ thin films on *c*-plane sapphire substrates have an epitaxial monoclinic phase at room temperature. Supplementary Fig. S1b shows the *R-T* curve of the VO$_2$ thin film and the corresponding differential curve in the inset. The transition temperature for the insulator-metal-transition (IMT) is around 344 K. The change of resistance across the IMT is up to four orders of magnitude, consistent with literature reports[2]. The structural studies and electrical measurements demonstrating the IMT in the oriented films both highlight the representative nature of the sample.

**Rutherford backscattering spectrometry (RBS) of VO$_2$ films grown on Al$_2$O$_3$ substrates**
RBS was performed using a beam of 2.3 MeV 4 He$^{2+}$ ions on representative film grown under similar condition. In RBS, the high-energy projectiles backscatter from the sample and provide information about the elemental composition[3]. The scattering angle of the beam, beam diameter, and energy resolution of the detector is 163 °, 2 mm, and 18 keV respectively. The energy per channel is 1.38 keV/ch. The experimental data were analyzed using the SIMNRA program. In Supplementary Fig. S2, we present the RBS data for the VO$_2$ film grown on c-plane sapphire substrate. As shown, the experimental and simulated data are in good agreement. The composition of the film obtained in terms of areal density from the SIMNRA analysis shows a high-quality stoichiometric film with V:O = 1:2.

**Supplementary Fig. S1**

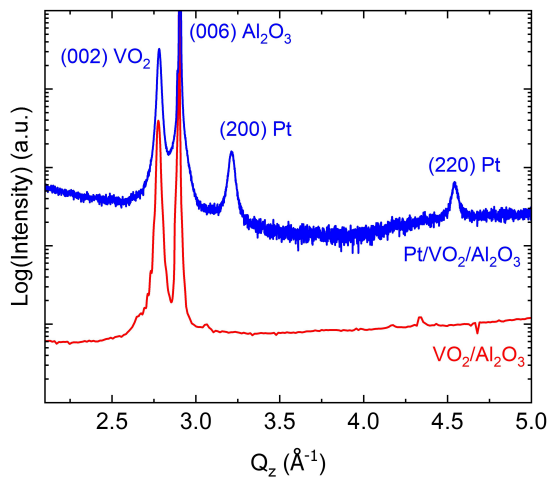 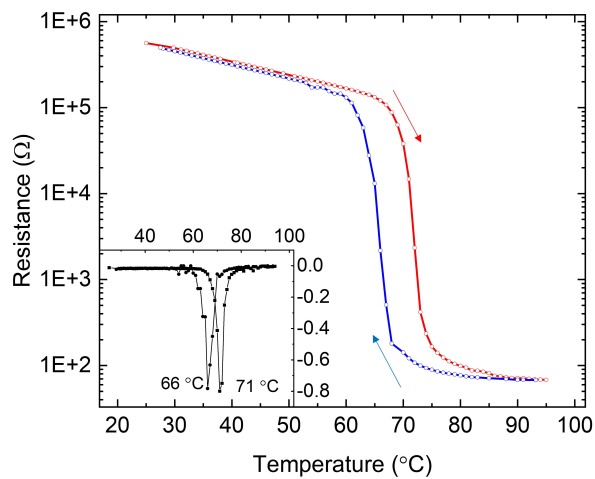

**Supplementary Fig. S1. Structural and electrical characterization of VO₂ devices.**
**a**, The XRD profile for VO$_2$/Al$_2$O$_3$ and Pt/VO$_2$/Al$_2$O$_3$ samples. **b**, Electrical resistance of a VO$_2$ film (100-nm thickness) as a function of temperature, measured with in-plane configuration. The red and blue arrows mark the heating and cooling curves respectively. The inset shows the derivative of the resistance as a function of temperature.

**Supplementary Fig. S2.**

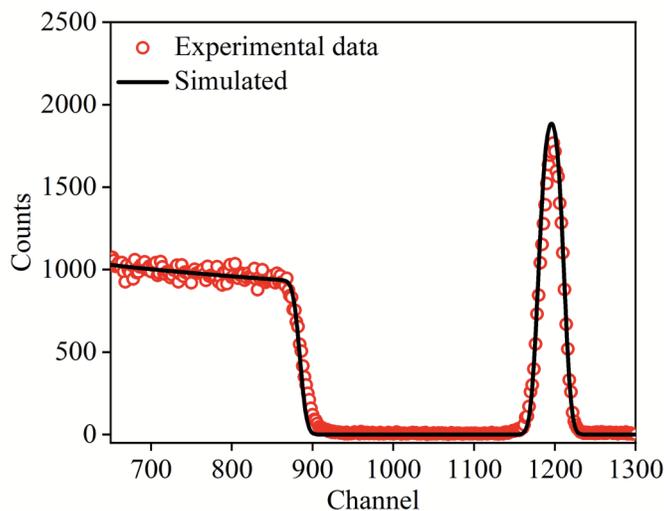

**Supplementary Fig. S2. Rutherford backscattering spectrometry of VO₂ films.** RBS spectra for epitaxial VO$_2$ film grown on c-plane sapphire substrate. The dotted and solid line represents experimental and simulated spectra respectively for nominally stoichiometric film with V:O ratio of 1:2.

# Supplementary Fig. S3

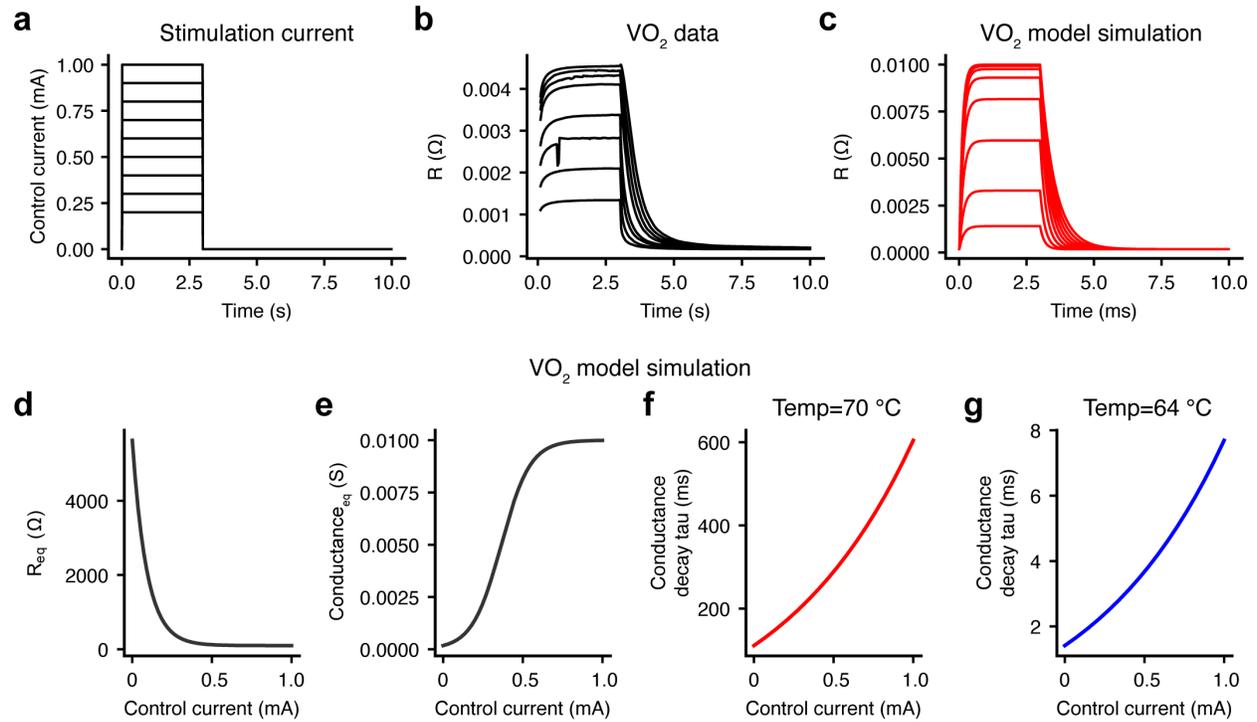

**Supplementary Fig. S3. Computational model of VO₂ device properties. a-c**, A computational model was constructed to fit the hardware measurements from a VO$_2$ device (70 °C) (**b**) in response to a range of applied current amplitudes (**a**). Data produced by the simulation (**c**) matches the dynamic range of conductance states, and the timescales of activation and relaxation (compare **b** and **c**). **d-e**, Model simulation data. The relationship between applied current and steady-state resistance (**d**) or conductance (**e**) is shown. **f-g**, The relationship between applied current and the time constant of conductance decay is shown with model temperature set to 70 °C (**f**) or 64 °C (**g**).

**Supplementary Fig. S4**

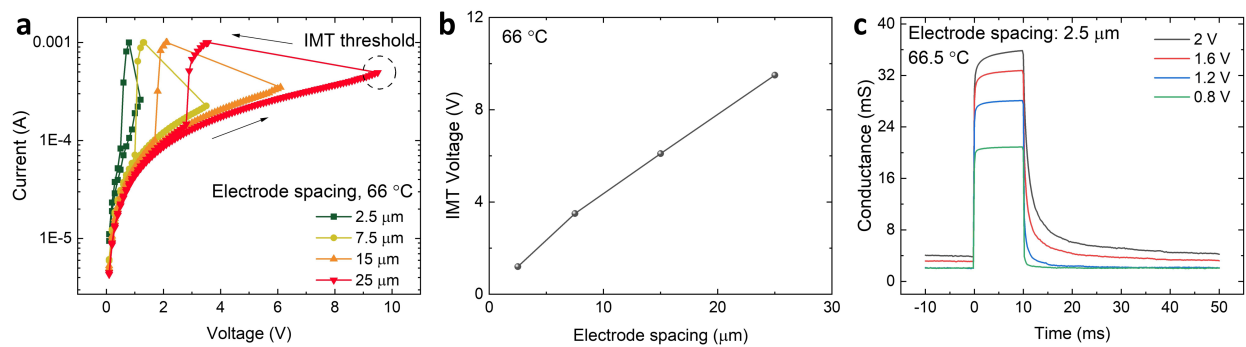

**Supplementary Fig. S4. Hardware measurements from smaller scale VO$_2$ devices.**
**a**, I-V measurements were performed on VO$_2$ devices with different electrode spacing at 66 °C. The compliance current was set to 1 mA. The schematic of the device is shown in Fig. 1a. As the channel spacing decreases, the insulator-metal transition (IMT) threshold voltage decreases, a trend that is consistent with the literature[4]. **b**, The IMT voltage as a function of electrode spacing. Data are extracted from panel **a**. **c**, Brief voltage pulses (10 ms) with varying amplitude (0.8 - 2 V) were applied to a 2.5 µm device, demonstrating accessibility of intermediate conductance states and slow phase relaxations (>10 ms) at a lower voltage than the 100 µm device characterized in Fig. 2.

**Supplementary Fig. S5**

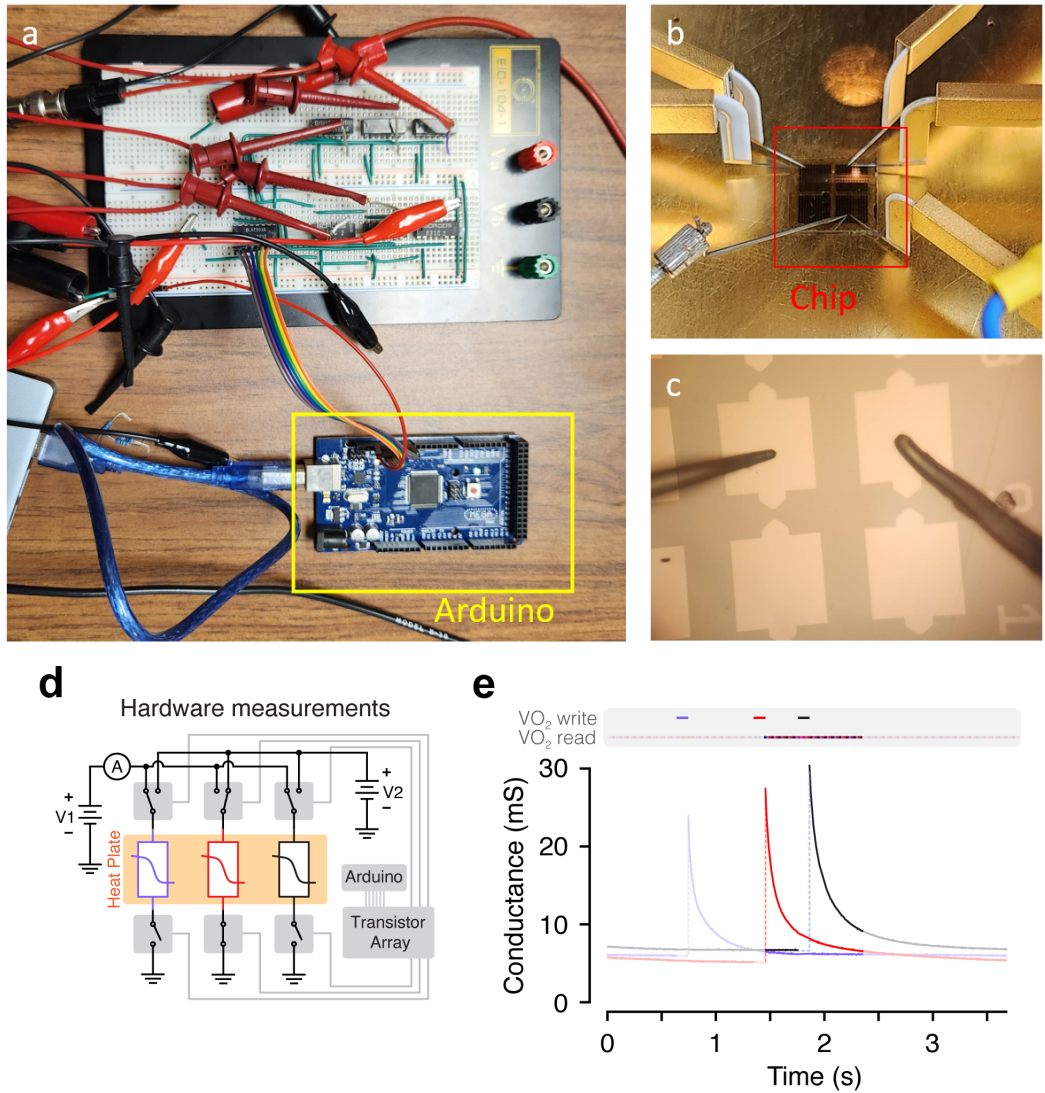

**Supplementary Fig. S5. Experimental setup for hardware measurements.**
**a**, Photograph of experimental setup. **b**, Six electrical probes are connected to three $VO_2$ devices. **c**, High magnification view of an in-plane $VO_2$ device. **d**, Diagram shows experimental setup used to perform physical hardware measurements from multiple neighboring devices in a $VO_2$ array. Three $VO_2$ devices were connected to a source of current to control and record their variable resistances. A digital control circuit (Arduino) was used to control a set of switches to control the timing of stimulation of the $VO_2$ devices. **e**, Hardware measurements demonstrate that the conductances of three $VO_2$ devices with similar relaxation timescales can be independently controlled and recorded. Read pulses were alternated between 3 devices for the duration of the experiment. The non-shaded region highlights a subset of read pulses that would be required when emulating biochemical signals according to the synaptic plasticity model shown in Fig. 4.

## Supplementary Fig. S6

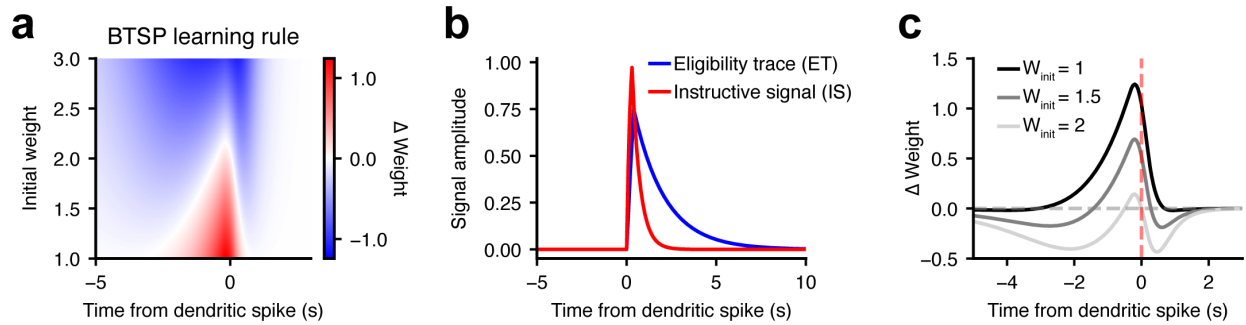

**Supplementary Fig. S6. Behavioral timescale synaptic plasticity (BTSP) learning rule. a**, Given a dendritic calcium spike in a postsynaptic neuron, any presynaptic input that is active within a seconds-long time window surrounding the dendritic spike will undergo a change in synaptic weight that depends on the timing of its activation and its initial synaptic weight[5]. Here it was assumed that a dendritic spike lasted for 300 ms, and that a single presynaptic input was activated for 400 ms. The relative timing between the onset of presynaptic activity and the onset of a postsynaptic dendritic spike was varied, and the mathematical model of the BTSP learning rule was used to predict the resulting changes in synaptic weight[5] (see Methods). The heatmap shows the change in weight produced by a single pairing with a dendritic spike for a range of activation times (x-axis) and initial weights (y-axis). **b**, Biochemical signals required for BTSP were generated by a computational model of a $VO_2$ device. Presynaptic activity generated a slowly decaying synaptic eligibility trace (ET, blue), and the postsynaptic dendritic spike generated a slowly decaying instructive signal (IS, red). **c**, The timescale of the learning rule and the balance of synaptic potentiation and depression is shown for three values of initial synaptic weight.